\documentclass[9pt,academicons]{article}

\usepackage{CADA}
\usepackage{comment}
\usepackage{booktabs}
\usepackage{makecell}
\usepackage{enumitem}

\title{Transparency Rendering in Computer-Aided Design:\\
Methodologies, Trade-offs, and Challenges}

\begin{document}

\maketitle

\authorSection{
    \anAuthor{Grigoris Tsopouridis}{0000-0001-8033-5481}{1},
	\anAuthor{Ioannis Fudos}{0000-0002-4137-0986}{1}
}

\affiliationSection{
	\anAffiliation{1}{University of Ioannina}{\{g.tsopouridis,fudos\}@uoi.gr}
}


\correspondingAuthor{Ioannis Fudos}{fudos@uoi.gr}


\abstract{
This paper surveys the state of transparency rendering in Computer-Aided Design (CAD), with a focus on both practical deployment in industrial systems and the underlying algorithms. We first review current approaches to transparency rendering in CAD environments and outline application scenarios in which accurate and performant transparency is critical for design inspection, communication, and decision-making. We then analyze the trade-offs between approximate and exact transparency techniques, comparing their performance-quality balance on desktop and mobile platforms and discussing criteria for selecting appropriate methods. The survey further identifies the need for robust benchmarks, quality metrics, and evaluation methodologies tailored to CAD-specific visualization tasks. We examine techniques for 
emphasizing important interior components such as importance-driven transparency, silhouette-based methods, and related approaches to support effective spatial understanding in complex 
assemblies. Finally, we discuss the unique challenges of rendering transparent constructive solid geometry (CSG) objects, including robustness, correctness, and integration with modern rendering pipelines. Collectively, these contributions characterize current capabilities, systematize open problems, and outline future research directions for transparency rendering in CAD visualization.
}

%

\doi{10.14733/cadaps.2027.133-151}


\section{INTRODUCTION}
\begin{figure}[!h]
    \centering
    \includegraphics[width=1\linewidth]{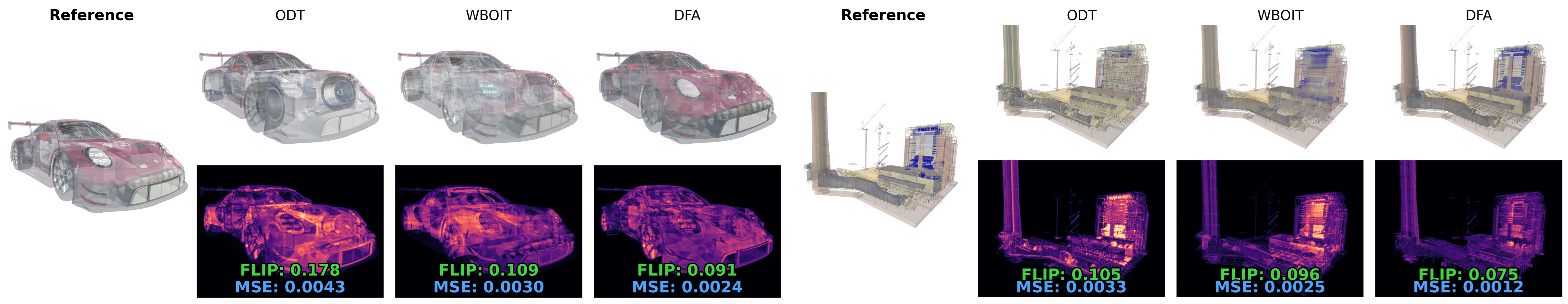}
    \caption{Transparency rendering on the Porsche and Powerplant CAD scenes. Top: A-Buffer reference, ODT, WBOIT, and DFA. Bottom: FLIP \cite{flip-paper} error maps with FLIP mean and MSE (lower is better). DFA achieves the lowest error, while ODT loses the most detail.}
    \label{fig:qualityhorizontal}
\end{figure}
Transparency rendering in Computer-Aided Design (CAD) must balance two requirements that are rarely satisfied together: pixel-accurate compositing of many overlapping, semi-transparent parts and interactive performance on a wide spectrum of hardware. In CAD review and inspection, transparency is not merely aesthetic; it conveys spatial relationships, clearances, and interferences inside complex assemblies where depth complexity can be extreme. Traditional order-dependent approaches or naive geometry sorting fail under such conditions, while exact Order-Independent Transparency (OIT) techniques can become prohibitively expensive in memory and bandwidth as depth complexity grows \cite{Carpenter1984,Everitt2001}. Approximate and hybrid OIT reduce cost but may compromise edge fidelity, silhouettes, and the perceived correctness of depth ordering, especially in high-opacity regimes that are common in engineering inspection \cite{McGuire2013,Munstermann2018}. The practical problem is to select, configure, and, when appropriate, augment transparency methods so that engineers consistently see the right structure at the right time on desktop workstations, legacy devices, and mobile/XR platforms-without breaking interactivity or trust.

This paper makes four contributions. First, it systematizes the design space of transparency rendering for CAD by comparing exact, approximate, and hybrid OIT families, including recent neural variants, through the lens of CAD-specific constraints (precision, topology, very high instance counts, and large coordinate ranges) and goals (clarity, stability, interpretability) \cite{Maule2013,Tsopouridis2022,tsopouridis2024dfa}. Second, it presents a performance, memory, and quality analysis that highlights where exact methods become infeasible, where blended or moment-based methods are sufficient, and when hybrid or neural approaches offer the best balance across desktop and mobile platforms. Third, it introduces visibility enhancements tailored to CAD workflows-significance-driven emphasis, silhouette-based highlighting with transmittance-aware attenuation, and z-fighting-robust ordering via a packed depth key-showing how these orthogonally improve legibility regardless of the underlying OIT pipeline. Fourth, it outlines practical guidance and selection criteria for platform-aware deployment, discusses challenges unique to CAD such as transparent CSG robustness and evaluation metrics focused on edges and structural integrity, and identifies open problems and promising directions for future research.

The remainder of the paper is organized as follows. In the remainder of this Section we begin by reviewing core transparency rendering concepts and OIT families relevant to CAD, along with background on CAD geometry representations and visualization-oriented meshing that condition rendering inputs. In Section \ref{sec:comparison}, we provide a comparison of rendering for CAD versus graphics/entertainment to ground the differing priorities that shape technique choice. Next, in Section \ref{sec:motivation} we motivate why fast, accurate transparency is pivotal in CAD and survey exact, approximate, hybrid, and neural methods with attention to their trade-offs. Section \ref{sec:performance} presents an empirical analysis of performance and memory across representative scenes and platforms, and a discussion of quality considerations most salient to engineering review. In Section \ref{sec:important-parts}, we introduce CAD-focused visibility enhancements-importance-driven opacity, silhouette cues with transmittance attenuation, and strategies to mitigate z-fighting-and show how they compose with different OIT pipelines. Finally, Section \ref{sec:conclusions} discusses challenges, including CSG transparency and CAD-specific metrics, presents practical recommendations for desktop and mobile/XR deployment, and concludes with avenues for future work.

\subsection{Transparency Rendering}
\label{sec:sota-transparency-rendering}
Transparency rendering aims to correctly display scenes that contain translucent or transparent objects, such as glass, plastic, or semi-transparent parts of an assembly. The core difficulty is that the final color at each pixel depends on the contributions of \emph{all} overlapping fragments along the viewing ray, combined in a specific front-to-back order. Traditional approaches often rely on sorting geometry before drawing, or on approximations that implicitly assume a simple depth ordering. In complex CAD models with many intersecting parts and high depth complexity, these assumptions break down, leading to visible artifacts such as incorrect blending or popping when the view changes.

Order-Independent Transparency (OIT) removes the dependency on geometry draw order by explicitly managing the fragments that contribute to each pixel. Exact OIT methods, such as depth peeling~\cite{Everitt2001,Bavoil2008OrderIT}, A-buffer techniques~\cite{Carpenter1984,yang2010,Vasilakis2012}, and k-/k$^{+}$-buffer variants~\cite{kbuffer,kplusbuffer}, store multiple fragments per pixel, sort them by depth, and then blend them in the correct order. These approaches provide high visual accuracy but require significant memory and bandwidth, and their cost grows quickly with depth complexity~\cite{maule2011}, which can limit their use in large, detailed CAD assemblies or on mobile hardware.

To reduce cost, many \emph{approximate} OIT techniques have been proposed. Blended methods such as Weighted Average Transparency (WAVG)~\cite{Bavoil2008OrderIT} and Weighted Blended OIT (WBOIT)~\cite{McGuire2013} avoid explicit per-pixel fragment lists and instead accumulate colors and opacities using carefully chosen blending formulas. They are simple to integrate into existing rendering pipelines and are efficient in both computation and memory, but can produce noticeable errors in scenes with high opacity or strong depth layering. Moment-Based Transparency (MBOIT)~\cite{Munstermann2018} improves the fidelity of blended methods by storing a small set of statistical moments per pixel and reconstructing transmittance from these moments, at the cost of additional computation and more complex shading. Hybrid methods such as Hybrid Transparency (HT)~\cite{Maule2013,Tsopouridis2022} and Multi-Layer Alpha Blending~\cite{Salvi2014} combine a limited exact representation (e.g., a small k-buffer for the nearest layers) with a cheaper approximate treatment for deeper fragments, offering an explicit performance–quality trade-off.

Recent work explores the use of neural networks to further improve this trade-off. Although most neural rendering research has focused on problems such as view synthesis or shadows~\cite{Datta2022}, similar ideas can be applied to transparency. Deep Hybrid OIT~\cite{Tsopouridis2022} augments variable k-buffer methods with a neural network that predicts which fragments are most important visually, enabling better quality under the same memory budget. Neural Moment Transparency~\cite{Tsopouridis2024NeuralMT} builds on MBOIT by using a neural network to recover transmittance more accurately from moment data. Deep and Fast Approximate OIT (DFAOIT)~\cite{tsopouridis2024dfa} goes one step further and directly predicts the final transparent color per pixel from simple per-pixel statistics (such as average color and opacity), providing a highly efficient solution suitable for real-time CAD visualization where responsiveness is critical.
Current commercial CAD systems use offline ray tracing for high quality rendering. For real-time transparency in CAD, vendors typically combine techniques: exact or semi-exact OIT (e.g., A-buffer or k-buffer), hybrid schemes, and weighted-blended OIT (e.g., WBOIT). Implementation details vary by product and version-AutoCAD and Rhino are reported to offer multi-layer (A-/k-buffer–like) modes in some pipelines, while BETA CAE Systems provides hybrid/weighted-blended options. When OIT is unsupported or too costly on a given GPU, applications generally fall back to screen-door or x-ray style transparency.

\subsection{CAD Model Representations and Meshing for Visualization}
\label{sec:cad-models}
Modern CAD systems represent product geometry using boundary representations (B-reps) built from analytic surfaces and free-form spline patches, organized in a topological data structure of faces, edges, and vertices~\cite{hoffmann1989geometric,mortenson2006geometric}. Free-form portions of a model are typically encoded with Non-Uniform Rational B-Splines (NURBS), which provide a unified framework for representing curves, surfaces, and many standard analytic primitives with high accuracy and compactness~\cite{piegl1997nurbs,farin2002curves}. In addition to the geometric description, CAD models carry rich semantic information (feature history, design intent, constraints, tolerances, assemblies and kinematic relationships), which is crucial for engineering workflows but largely invisible in conventional rendering pipelines~\cite{shah1995parametric,brepFeature2001}. For visualization, however, the core input remains the precise geometric and topological description of the 3D boundary, often specified in a product data exchange format such as STEP or IGES~\cite{iso10303,igescitation}.

To leverage commodity graphics hardware, these continuous and often curved CAD surfaces are converted into polygonal meshes through tessellation or meshing algorithms~\cite{frey1999mesh,owen1998survey}. Visualization-oriented meshing aims to approximate the exact geometry with controlled deviation while keeping the mesh complexity manageable, often using adaptive refinement that accounts for curvature, view-dependent error, and application-specific tolerances~\cite{botsch2010polygon,garland1997surface}. Many industrial systems perform on-the-fly tessellation of NURBS patches, generating triangle meshes that satisfy chordal- and normal-deviation criteria, followed by optional mesh optimization such as smoothing or decimation~\cite{schroeder1992decimation,shewchuk1996quality}. The resulting polygonal representation is then used by the rendering pipeline for shading, hidden-surface removal, and transparency computations, while the exact CAD model remains available for downstream operations such as editing, simulation, and manufacturing.

\section{RENDERING FOR CAD VS RENDERING FOR GRAPHICS}
\label{sec:comparison}

Computer-Aided Design (CAD) rendering and graphics/entertainment rendering share a common foundation in rasterization, shading, and GPU acceleration, yet they optimize for different outcomes. CAD serves engineering decision-making where correctness, stability, and interpretability drive value. Graphics for entertainment prioritizes visual plausibility, stylistic control, and throughput, where what the audience perceives as real or engaging is paramount. This section presents a unified description of their differences and common ground, elaborating on goals, geometry, precision, scene complexity, shading and transparency, interaction and lighting, robustness and system integration, and the specific importance of fast, accurate transparency for CAD workflows.

\subsection{Goals and Geometry Representation}
In CAD, the primary goal is accurate understanding of exact geometry in support of design and inspection. Visualization is instrumental rather than an end in itself: it must reliably reveal shapes, sizes, fits, and relationships so that engineers can reason about tolerances and manufacturability. Consequently, CAD models are expressed as exact mathematical entities-NURBS, boundary representations (B-reps), and analytic primitives. Assemblies are frequently enormous and deeply hierarchical, reflecting real products composed of many repeated parts and subassemblies, each bound by strict tolerances that affect downstream processes.

In graphics and entertainment, the goal is to achieve visual realism, style, and performance within the constraints of real-time or offline pipelines. Fidelity is evaluated by audience perception: if it looks right, it is right. Geometry is therefore represented as triangle meshes that approximate shape at a level tailored to the camera, the shot, and GPU budgets. Topology and polygon counts are aggressively optimized to maintain stable frame times and predictable memory footprints, and the mesh representation integrates cleanly with toolchains for baking, LOD generation, and deformation.

\subsection{Precision and Scene Complexity}
CAD workflows demand double precision and strict adherence to topological correctness. Even minute gaps, overlaps, or non-manifold conditions can invalidate analyses, lead to incorrect inferences during inspection, or create vulnerabilities in manufacturing. Although CAD scenes may use a relatively small variety of materials, they often exhibit extremely high instance counts, large coordinate scales, and repetitive part structures that strain instancing, culling, and precision management in rendering.

Graphics pipelines typically rely on 32-bit floating point precision, with continuity judged by visual plausibility. Minor cracks, slight overlaps, or non-manifolds are tolerated if they remain invisible under expected lighting and camera motion. Scenes emphasize asset diversity, material richness, post effects, and detailed shading models. Despite the complexity of materials, per-frame total part counts are often lower than in CAD, aided by content authoring that targets performance budgets.

\subsection{Shading, Materials, and Transparency}
CAD favors simple, robust material models that communicate form and function clearly. Flat shading, Phong, basic PBR, and false-color overlays support interpretability and analysis. Edges, silhouettes, cross-sections, and display modes such as wireframe, hidden-line removal, and sectional views are central to reading design intent. Transparency is often implemented through predictable, conservative techniques (e.g., x-ray, screen-door patterns, or simple alpha blending) that balance clarity and speed over perfectly accurate order-independent results. The objective is to reveal structure reliably, not to create cinematic translucency.

Graphics and entertainment rendering invests heavily in complex PBR models with layered BRDFs, subsurface scattering, volumetrics, and post-processing pipelines that yield cohesive artistic results. Transparency leverages sophisticated methods such as order-independent transparency (OIT), screen-space reflections (SSR), and volumetric fog, selected to support mood, depth cues, and scene coherence. The bar is ``cinematic'': physically based where helpful, stylized where desired, always in service of the final image.

\subsection{Why Fast, Accurate Transparency Matters in CAD}
Modern CAD assemblies can contain thousands of overlapping parts. Engineers routinely make outer shells or covers semi-transparent to inspect interior geometry without losing spatial context. Achieving this requires two properties simultaneously: accuracy, so that the per-pixel blending order respects true depth relationships, and performance, so that interactive frame rates are maintained even for large, complex assemblies. If either is compromised, the workflow suffers: a slow view breaks the inspect–adjust–reinspect loop; a visually fast but incorrect blend erodes trust and risks misjudged geometry.

The cost of getting transparency wrong extends beyond aesthetics. Incorrect blending can invert depth perception, hiding interferences or inventing clearances that do not exist. Latency disrupts decision-making in review sessions where multiple stakeholders rely on a single model, whether on a workstation display or in XR. Downstream, visualization mistakes that persist into manufacturing or on-site work become far more expensive to correct than those caught during design review.

Transparency also underpins key applications: design inspection (confirming fit and placement; spotting clashes; tracing cables and hoses through enclosures), design communication and review (maintaining context while highlighting subsystems; sustaining confidence on shared screens or in VR/AR), and real-time decision-making (overlaying simulation fields like stress or thermal data beneath translucent shells; ensuring that model edits appear both correct and responsive before the next decision). In each case, accuracy and performance are not optional embellishments-they are the conditions for reliable engineering judgment.

\subsection{Similarities and Convergences}
Despite differing priorities, CAD and graphics share core techniques: both rely on GPU pipelines, mesh processing, culling, and shading models; both balance quality against frame time; both use transparency to reveal otherwise hidden structure; and both benefit from robust anti-aliasing, stable temporal behavior, and readable lighting. There is growing convergence: basic PBR materials can improve the legibility of CAD surfaces; engineering viewers increasingly adopt lightweight OIT variants for clarity; graphics pipelines borrow analytical views and outlines for gameplay readability; and both domains integrate with XR, demanding high frame rates and predictable latency. The shared challenge is selecting techniques that maximize human understanding for the task at hand.

Rendering for CAD and for graphics/entertainment diverge because they answer to different definitions of success. CAD prioritizes correctness, stability, and interpretability over stylistic flourish, operating at extreme scene scales with exact geometry, strict topology, and deep integration with engineering data. Graphics prioritizes perceptual realism, artistic intent, and sustained performance under motion, leveraging complex materials, lighting, and effects tuned to audience perception. Transparency exemplifies the split: CAD needs fast and unequivocally correct blending to support decisions; graphics seeks convincing and expressive translucency that serves composition. Even so, the two domains are not opposites. They share tools, evolve under common hardware constraints, and increasingly borrow techniques where they add clarity without sacrificing core objectives. The best rendering is the one that most effectively communicates truth to its audience-engineer or player-within the time available to decide what comes next.

\section{TRANSPARENCY FOR COMPUTER AIDED DESIGN}
\label{sec:motivation}
Transparency is not a cosmetic feature in CAD visualization: it is a primary method of interaction. Engineers use it to inspect internal components without suppressing an assembly's context, to verify clearances and interferences between nested parts, and to communicate design intent to non-expert stakeholders during reviews. Unlike entertainment applications, where an incorrect blend mostly affects aesthetics, an incorrect transparency computation in CAD can misrepresent geometry that a decision depends on: a hidden interference, a missed edge, or a component silhouette that should have been visible but is lost in the blend. Two distinct failure modes motivate the trade-offs surveyed in this section: \textit{quality failures} and \textit{performance failures}. 
A quality failure occurs when the composited image is a poor approximation of the exact front-to-back blend (A-Buffer\cite{yang2010}), discarding or misordering fragments so that geometric detail disappears or appear inaccurately. Figure \ref{fig:qualityhorizontal} shows a representative case where the correctly rendered surface retains the curves and details of the inside parts, while the same surfaces under some of the inexpensive approximations lose part of their internal structure entirely, precisely the kind of missing detail an inspection workflow cannot tolerate. A performance failure occurs when a technique reconstructs an accurate blend but at a memory or bandwidth cost that a target platform cannot sustain, a common outcome once the depth complexity of a full CAD assembly (potentially thousands of overlapping transparent parts) is combined with the resolution demands of modern design-review displays. The performance failures are especially important for mobile, used for AR inspection, and legacy devices. As we show in Section 3, exact methods historically incur the highest cost in exactly this regime, while approximate methods trade away detail preservation for bounded, resolution-independent cost. Because CAD assemblies routinely combine both high depth complexity and a demand for reliable detail, no single category dominates, which is why CAD systems in practice mix categories rather than committing to one (as noted for AutoCAD, BETA CAE, and Rhino in Section \ref{sec:sota-transparency-rendering}).

Transparency methods have converged on a small number of recurring strategies for this problem, which we group here into three categories: exact approaches that reconstruct the correct per-pixel blend without approximation, approximate approaches that trade accuracy for bounded, predictable cost, and hybrid approaches that combine the two, typically by treating only a small subset of the geometry exactly.

\subsection{Exact Approaches}
Exact approaches guarantee that the blending is evaluated correctly at every pixel: every fragment along a viewing ray is captured, sorted by depth, and composited in the right order. The original solution to this problem, the A-buffer~\cite{Carpenter1984}, stores every incoming fragment in a per-pixel data structure, either as a fixed-capacity array indexed per pixel or as a dynamically grown linked list~\cite{yang2010}, leading to unbounded memory requirements, which is especially severe for large CAD assemblies. Depth  peeling~\cite{Everitt2001} reformulates the same problem as a sequence of rendering passes, each of which extracts exactly one depth layer, avoiding an explicit per-pixel list or array, and bounding the memory requirements, at the cost of re-rasterizing the scene once per layer, making this method unsuitable for real-time applications due to its high performance requirements. The $k$-buffer and its synchronized variants~\cite{kbuffer,kplusbuffer} narrow this further by retaining only a bounded number of nearest \textit{k} fragments per pixel, which is exact whenever the true depth complexity does not exceed the buffer's capacity, and becomes an approximation otherwise.

A fixed $k$ across the entire image, however, allocates the same layer budget to every pixel
regardless of how much depth complexity that pixel actually has, wasting capacity in simple
regions while starving complex ones. Vasilakis et al.~\cite{Vasilakis2017}
address this with a variable $k$-buffer that assigns a per-pixel layer budget $k(p)$ according to an importance map estimated dynamically per frame, allocating more layers to regions that
contribute more to the final image and reclaiming capacity from regions that do not. For a
fixed overall memory budget this yields a better distribution of exactness across the image
than a uniform $k$, which is directly relevant to CAD assemblies: depth complexity in a typical
scene is highly non-uniform, concentrated around nested or clustered components, and a
per-pixel budget can devote more of the available exact layers to precisely those regions
without increasing total memory cost.

For CAD visualization specifically, exact methods are attractive because they are the only
category that does not, by construction, risk any kind of silhouette or detail or interior-structure loss as every fragment that should contribute to a pixel does. The cost, as
Table \ref{tab:memory} and Table \ref{tab:memory-example} make explicit, is memory that scales with the actual depth complexity $n(p)$ of the scene rather than with a fixed budget.

\subsection{Approximate Methods}
Approximate methods abandon the guarantee of an explicit per-pixel fragment list and instead
accumulate a small, fixed set of statistics as each fragment is rasterized, reconstructing an
estimate of the correct blend from those statistics alone, usually using a modified blending equation. Weighted Blended OIT (WBOIT)~\cite{McGuire2013} is the most widely deployed example: each fragment contributes to a running color and alpha accumulator weighted by a function of its opacity and depth, so that the entire technique requires only one extra color buffer and one scalar channel regardless of scene depth complexity. Moment-Based OIT (MBOIT)~\cite{Munstermann2018,Sharpe2018} improves on this by replacing WBOITs hand-tuned weighting function with a small number of statistical moments of the depth distribution, from which transmittance is reconstructed analytically, at roughly double the memory cost of WBOIT for a comparable configuration, however this method features two geometry passes, vastly increasing computational demands. Neural Moment Transparency \cite{Tsopouridis2024NeuralMT} improves the reconstruction quality by utilizing a neural network for transmittance computation.

Recent work replaces these hand-designed accumulation formulas with a learned one. Deep and
Fast Approximate OIT (DFAOIT)~\cite{tsopouridis2024dfa} keeps only two exactly-tracked
fragments plus a compact set of per-pixel statistics (average color, average opacity,
accumulated premultiplied color) and feeds them to a small neural network trained to predict
the entire blended result directly, in place of a hand-derived weighting or moment 
reconstruction formula. Despite touching no more per-pixel state than WBOIT or MBOIT
(Table~\ref{tab:memory}), it achieves markedly better accuracy than either, and does so without the
scene- and camera-specific tuning that WBOITs weighting function demands, which makes it a
better drop-in candidate for CAD viewers that must render arbitrary, previously unseen
assemblies without per-scene calibration. This method utilizes one to three geometry passes and an inference pass, having a small performance overhead but higher quality compared to other approximate methods. A recent work, STAR-NT \cite{tsopouridis2026starntspatiotemporalaccelerationrealtime} optimizes DFAOIT performance by adaptively setting each geometry pass's resolution, with a small quality degradation. 

The appeal of this category for CAD applications is precisely the inverse of the exact methods limitation: memory and bandwidth cost is bounded and independent of assembly complexity, which is what makes real-time transparency feasible at all on constrained hardware, including the mobile and legacy hardware. The cost is quality, and here CAD imposes a stricter standard than most entertainment based applications as even the best of these methods are prone to under or over-representing the nearest, most important surfaces when overall opacity is high, precisely the regime in which a designer is most likely to be inspecting a specific highlighted part through several layers of surrounding, mostly-transparent geometry. This tension between bounded cost and detail preservation is the central trade-off that importance-driven and silhouette-based enhancement techniques we present (Section \ref{sec:important-parts}), since they can restore emphasis on a specific part without requiring the underlying transparency method itself to become exact.

\subsection{Hybrid Methods}
Hybrid methods split the fragments at a pixel into two groups and treat each with a different
strategy. A usually small number of nearest layers are composited exactly, typically via a bounded $k$-buffer, while the remaining, more distant fragments (the \emph{tail}) are blended using an approximate method. Hybrid Transparency~\cite{Maule2013} composites the $k$ closest fragments exactly and folds everything beyond that into a weighted-average-style \cite{Bavoil2008OrderIT} blended tail, so that memory cost grows with $k$ rather than with the true depth complexity $n(p)$, while still preserving exact detail for the layers nearest the camera and taking into account an approximate contribution of the remaining layers. Deep Hybrid OIT~\cite{Tsopouridis2022} retains a variable $k$-buffer structure, meaning that it has a different k value for each pixel \cite{Vasilakis2017}, but uses a neural network to decide, per pixel, which pixels are important enough to allocate more memory to, improving quality for the same memory demands.

For CAD, hybrid methods offer a direct compromise as they offer exact compositing near the camera, while a bounded memory footprint keeps performance predictable. An open question, revisited in Section \ref{sec:important-parts}, is how to decide \emph{which} fragments deserve exact treatment when that choice should reflect design semantics rather than depth order alone, as CAD rendering has different aims compared to accurate transparency rendering for entertainment applications.

\section{PERFORMANCE/QUALITY TRADEOFFS}
\label{sec:performance}

This section compares performance, memory requirements, and quality, and presents trade-offs important for selecting an appropriate method for a CAD system or platform.

\subsection{Performance}
\begin{figure}[!h]
    \centering
    \includegraphics[width=0.85\linewidth]{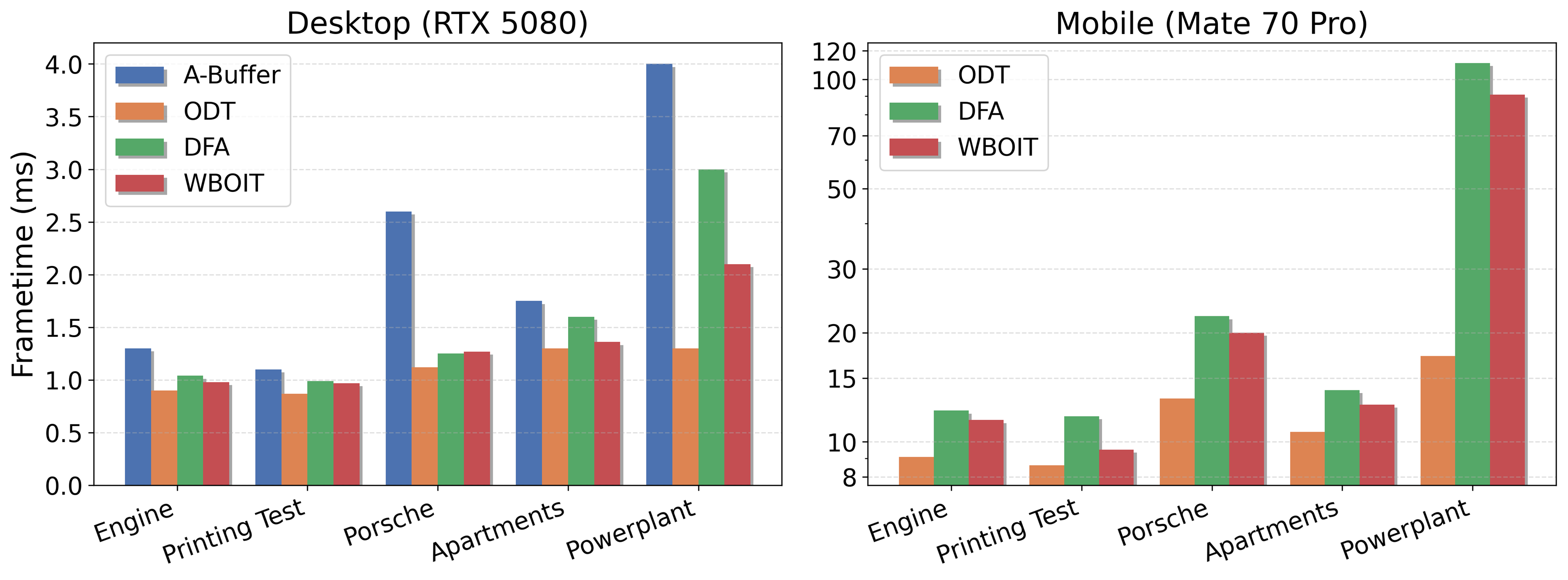}
    \caption{Measured frametime (ms) of representative transparency methods across five CAD
scenes of increasing depth complexity, on a desktop workstation (NVIDIA RTX 5080, linear scale)
and a mobile device (Huawei Mate 70 Pro). A-Buffer is omitted from the mobile plot
as it could not be executed within the device's memory and performance budget.}
    \label{fig:perfgraph}
\end{figure}

Table~\ref{tab:perf-combined}, and Figure \ref{fig:perfgraph} report measured frametimes across the same five scenes on two representative platforms: a desktop workstation with a discrete RTX 5080 GPU, and a mobile device (Huawei Mate 70 Pro) representative of increasingly common on-site, mobile-based review workflows, such as AR-based field inspection. The comparison includes an exact method (A-Buffer), the naive object-sorted painter's algorithm (ODT) as a performance
lower bound, a neural approximate method (DFA), and a blended approximate method (WBOIT).

On desktop hardware, all four methods remain within fast, real-time frametimes even for the most demanding scene (Powerplant), and the gap between the fastest (ODT) and slowest (A-Buffer) method is modest. This is the regime in which the accuracy versus cost trade-off discussed in Section \ref{sec:motivation} is, in practice, close to free: a desktop CAD workstation can afford an exact method without a noticeable performance penalty, and the choice of technique can be dictated by quality requirements alone. Most of those methods are expected to perform well on high-end workstations, however, lower-end and legacy devices may face performance hits similar to mobile platforms.

The mobile results tell a different story. A-Buffer could not be executed at all within the
device's memory and computational budget for any of the five scenes, confirming the memory-scaling argument empirically rather than only theoretically: exact per-fragment storage is not
merely slower on constrained hardware. Among the methods that do run, frametimes increase by roughly an order of magnitude relative to desktop across the board, which is expected given the difference in raw compute and bandwidth. More importantly, the increase is not uniform across methods. On the Powerplant scene, which is a very demanding CAD model, DFA and WBOIT are within 40\% of each other on desktop (3.00 vs.\ 2.10\,ms), but on mobile that gap widens sharply (111.11 vs.\ 90.91\,ms), and both fall well behind ODT (17.24\,ms) that exhibits sub-par accuracy. This scene, which has the highest depth complexity of the five, is precisely where the per-fragment overhead of DFAs neural inference and WBOITs weighting computation compounds most severely once compute is scarce, whereas ODTs cost is insensitive to depth complexity by construction, since it performs no per-fragment accumulation at all, only sorting sub-meshes by depth. For CAD systems that must support both desktop and mobile review, this suggests that method selection cannot be fixed once for an application: the same scene that renders comfortably under any of these methods on desktop may require falling back to a cheaper, lower-quality method on mobile, or restricting the more accurate methods to scenes below some depth-complexity threshold, an approach we revisit when discussing platform-aware method selection.

\begin{table}[h]
\centering
\caption{Frametime (ms) across representative CAD scenes of varying depth complexity, comparing
an exact method (A-Buffer), a naive object-sorted painter's algorithm (ODT), a neural
approximate method (DFA), and a blended approximate method (WBOIT), on a desktop
workstation (NVIDIA RTX 5080) and a mobile device (Huawei Mate 70 Pro). A-Buffer entries on
mobile are omitted (-) as the method could not be executed within the device's memory budget.}
\label{tab:perf-combined}
\begin{tabular}{l rrrr rrrr}
\hline
& \multicolumn{4}{c}{Desktop (RTX 5080)} & \multicolumn{4}{c}{Mobile (Mate 70 Pro)} \\
Scene & A-Buffer & ODT & DFA & WBOIT & A-Buffer & ODT & DFA & WBOIT \\
\hline
Engine        & 1.30 & 0.90 & 1.04 & 0.98 & - & 9.09  & 12.20  & 11.49 \\
Printing Test & 1.10 & 0.87 & 0.99 & 0.97 & - & 8.62  & 11.76  & 9.52  \\
Porsche       & 2.60 & 1.12 & 1.25 & 1.27 & - & 13.16 & 22.22  & 20.00 \\
Apartments    & 1.75 & 1.30 & 1.60 & 1.36 & - & 10.64 & 13.89  & 12.66 \\
Powerplant    & 4.00 & 1.30 & 3.00 & 2.10 & - & 17.24 & 111.11 & 90.91 \\
\hline
\end{tabular}
\end{table}

\subsection{Memory Requirements}
Frametime alone does not capture the full cost of a transparency method as memory is frequently also a binding constraint, especially on highly complex CAD designs featuring millions of triangles. We therefore treat memory as a first-class criterion alongside performance and quality, rather than as a footnote to either.

Table~\ref{tab:memory} summarizes the additional per-pixel memory requirement $m(p)$ of each
representative method, on top of the standard color and depth buffers already required by any
rasterization pipeline. We express $m(p)$ symbolically so that it can be evaluated for an
arbitrary configuration: $f$ denotes one RGBA color payload, $1$ denotes a single scalar
channel (depth, revealage, count, or a pointer), $k$ is a fixed layer capacity, $N$ is a
pre-allocated maximum layer count, $n(p)$ is the actual, scene-dependent depth complexity at
pixel $p$, and $b$ is the number of statistical moments retained by a moment-based method.
Methods with $s(p) = 0$ do not store individual fragments at all, relying instead on running
accumulators, which is precisely what keeps their cost independent of scene complexity.

\begin{table}[h]
\centering
\caption{The table summarizes the additional per-pixel memory requirements $m(p)$ of representative transparency methods, on top of the standard color and depth buffers. Here, $f$ denotes one RGBA color payload, $1$ denotes one scalar channel (such as depth, revealage, count, or a pointer), $k$ is the fixed layer capacity, $N$ is the pre-allocated maximum number of layers, $n(p)$ is the actual depth complexity at pixel $p$, and $b$ is the number of statistical moments. Methods with $s(p) = 0$ do not store individual fragments.}
\label{tab:memory}
\begin{tabular}{|l|l|c|l|}
\hline
\textbf{Method} & \textbf{Category} & $s(p)$ & $m(p)$ \\
\hline
Object-sorted                               & Object-sorted      & $0$    & $0$ \\
\hline
X-Ray                               & X-Ray      & $0$    & $0$ \\
\hline
Screen-door                                 & Pseudo-Transparent      & $0$    & $0$ \\
\hline
WBOIT~\cite{McGuire2013}                    & Approximate        & $0$    & $f + 1$ \\
\hline
MBOIT~\cite{Munstermann2018}                & Approximate        & $0$    & $\lceil b/4 \rceil \cdot f + 1$ \\
\hline
k-buffer~\cite{kbuffer}                     & Exact              & $k$    & $k \cdot (f + 1)$ \\
\hline
Hybrid Transparency~\cite{Maule2013}        & Hybrid             & $k$    & $k \cdot (f + 1) + f$ \\
\hline
A-Buffer (fixed array)~\cite{yang2010}      & Exact              & $N$    & $N \cdot (f + 1)$ \\
\hline
A-Buffer (linked list)~\cite{Carpenter1984} & Exact              & $n(p)$ & $n(p) \cdot (f + 2) + 1$ \\
\hline
DFAOIT~\cite{tsopouridis2024dfa}            & Neural approximate & $2$    & $4f + 3$ \\
\hline
\end{tabular}
\end{table}

To make these symbolic costs concrete, Table~\ref{tab:memory-example} evaluates $m(p)$ for a
representative configuration ($n(p) = 100$ on average, $f = 4$, $b = 8$, $k = 10$, $N = 300$)
and reports the resulting total additional memory at 4K resolution. The gap between categories
is stark: the bounded, accumulator-based methods (WBOIT, MBOIT, DFAOIT) stay within a few tens
to roughly 150MB regardless of scene complexity, while both A-Buffer variants exceed several
gigabytes, making them mostly unsuitable for CAD applications.

\begin{table}[h]
\centering
\caption{The table gives an example for the additional per-pixel memory requirements $m(p)$ of representative transparency methods, on top of the standard color and depth buffers. Here, $n(p)=100$ on the average, $f=4$, $b=8$, $k=10$, $N=300$.}
\label{tab:memory-example}
\begin{tabular}{|l|l|c|l|}
\hline
\textbf{Method} & \textbf{Category} & total for 4k resolution \\
\hline
Object-sorted                               & Object-sorted      & $0$  \\
\hline
X-Ray                               & X-Ray      & $0$  \\
\hline
Screen-door                                 & Pseudo-Transparent      & $0$    \\
\hline
WBOIT~\cite{McGuire2013}                    & Approximate        & $40\,MB$   \\
\hline
MBOIT~\cite{Munstermann2018}                & Approximate        & $71\,MB$    \\
\hline
k-buffer~\cite{kbuffer}                     & Exact              & $395\,MB$  \\
\hline
Hybrid Transparency~\cite{Maule2013}        & Hybrid             & $427\,MB$     \\
\hline
A-Buffer (fixed array)~\cite{yang2010}      & Exact              & $11,865\,MB$   \\
\hline
A-Buffer (linked list)~\cite{Carpenter1984} & Exact              & $4,754\,MB$ \\
\hline
DFAOIT~\cite{tsopouridis2024dfa}            & Neural approximate &  $150\,MB$  \\
\hline
\end{tabular}
\end{table}

Two practical consequences follow for a CAD system choosing among these methods. First,
memory cost for the semi-exact and hybrid categories is controlled by a user or system-chosen
parameter ($k$) rather than being fixed, which means the trade-off is tunable: reducing
$k$ recovers some of the memory needed, at the cost of reintroducing the risk of dropped fragments once $n(p)$ exceeds the chosen budget. Memory and frametime should therefore be evaluated as two separate axes of the same trade-off, not as proxies for one another, when selecting a method for a specific platform and scene.

\subsection{Quality}
We complement the previous two subsections with a quantitative and qualitative quality comparison across the same five CAD scenes and the same three practical methods (ODT, WBOIT, DFA), using the A-Buffer render as ground truth in every case. DFAOIT while a neural approximate, could be also partly considered as a hybrid method due to its ability to use the two closest fragments as an input feature, giving us an idea of where simpler hybrid methods stand in terms of quality.

\begin{figure}[h]
\centering
\includegraphics[width=\linewidth]{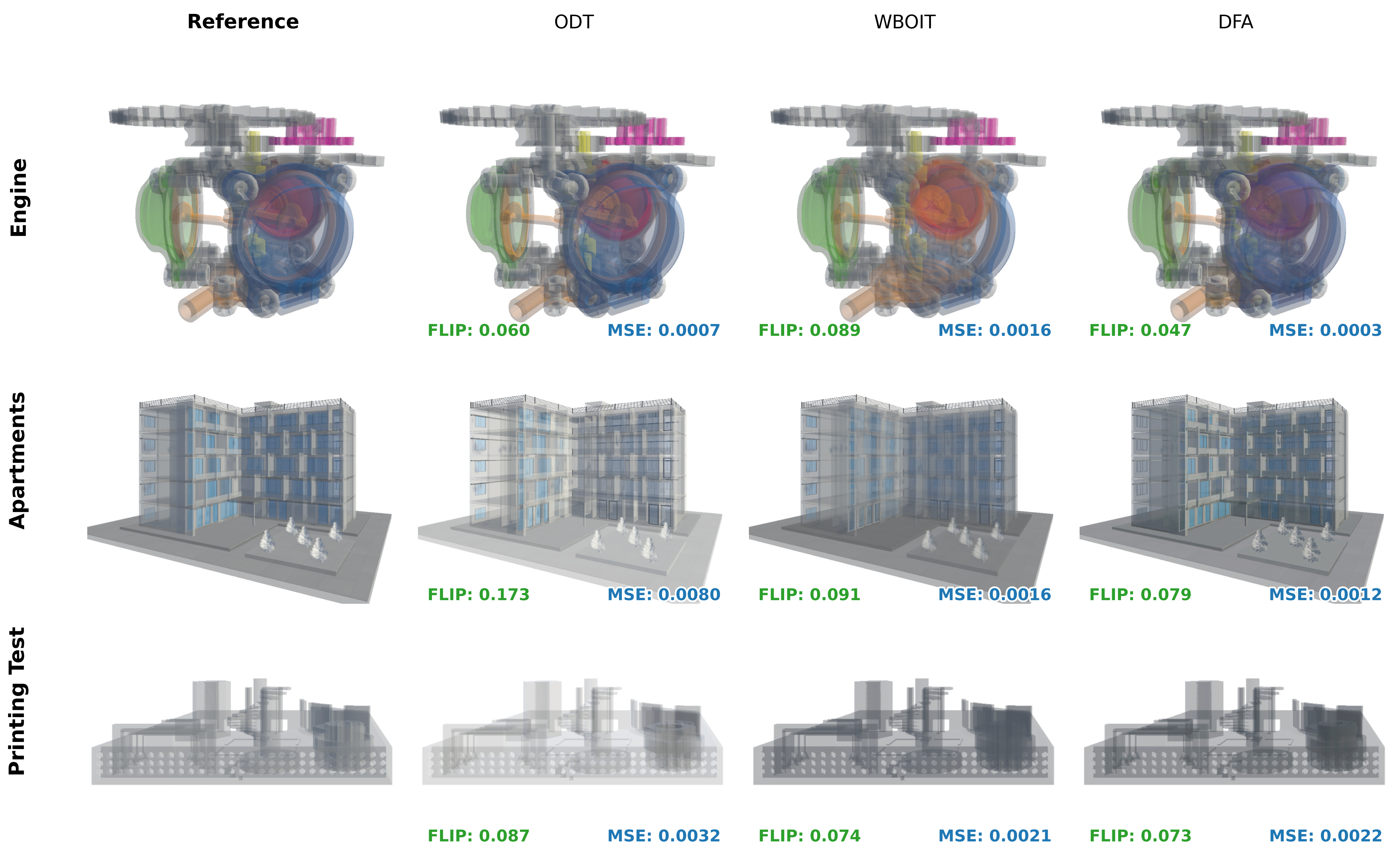}
\caption{Visual quality comparison across transparency methods for three additional CAD
scenes (Engine, Apartments, Printing Test), with the reference render shown first in each
row. Each method's render is annotated with its FLIP~\cite{flip-paper} mean score (green, lower is better) and MSE (blue, lower is better) relative to the reference.}
\label{fig:quality-comparison-2}
\end{figure}

We report two error measures per scene. Mean Squared Error (MSE) is included as a familiar,
easily interpreted baseline, but it is well known to correlate poorly with what a human
observer actually perceives as different, since it penalizes every pixel equally regardless of
whether the discrepancy falls on a flat, unremarkable region or on an important internal
edge that carries the geometric information CAD inspection depends on. We therefore report,
alongside MSE, the mean score under FLIP~\cite{flip-paper}, a perceptual difference metric
designed to approximate human sensitivity to exactly this kind of structured, edge-concentrated
error. Figures~\ref{fig:qualityhorizontal} and~\ref{fig:quality-comparison-2} show the corresponding renders and FLIP error maps.

Two patterns hold consistently across all five scenes. First, ODT is the worst performer on
both metrics in every single case, and by a wide margin on the three scenes with the most
intricate internal structure, Porsche, Powerplant and Apartments, where its FLIP score is roughly double that of WBOIT or DFA. This is the direct visual counterpart of the memory-free, simple object sorting cost profile noted in Section \ref{sec:performance}. Secondly, DFAOIT achieves the best on both metrics in most scenes, including the four scenes (Engine, Porsche, Apartments, Powerplant) where it strictly dominates WBOIT on both FLIP and MSE simultaneously. The one partial exception is Printing Test, which is an extremely simple scene.

The FLIP error maps in Figure ~\ref{fig:qualityhorizontal} make it clearer \emph{where} this error concentrates, which is at least as important for a CAD context as the aggregate score. Across every scene, error for all three methods is highest along dense, fine internal structure, exactly where Section \ref{sec:motivation} identified as most consequential for design inspection, rather than being spread uniformly across flat, unoccluded surfaces where any method reproduces the reference well. 

Taken together with the performance and memory results, this quality comparison points toward a clear default for CAD systems that must pick a single approximate method: DFA offers the best quality at a memory cost comparable to WBOIT (Table \ref{tab:memory}) and a frametime cost that remains acceptable on desktop and mobile hardware, making it the preferable approximate/hybrid choice whenever the target platform can absorb its higher per-fragment compute. ODT should be reserved for edge cases where it is genuinely necessary, the most depth-complex scenes on the most constrained mobile hardware, where its insensitivity to depth complexity is the only thing keeping the frametime interactive at all, and its quality cost should be treated as a deliberate, visible trade-off rather than a default. WBOIT remains a reasonable middle ground on hardware where DFAs inference cost is not affordable but ODTs quality loss is not acceptable either. In general, hybrid methods should offer quality similar or better compared to DFA, with increased computation and memory requirements. 

It is worth noting that both metrics used above, MSE and FLIP, were designed to answer a
question that is not quite the one a CAD user is \textit{always} asking. MSE quantifies raw color-space deviation, and FLIP improves on this by weighting error according to models of human contrast and color sensitivity, so that both ultimately optimize for pixels looking closer to the reference. This is the right goal for entertainment or photorealistic rendering, where the aim is a plausible or faithful \emph{image}. In CAD, the underlying goal is often different: what matters is not that a rendering looks close to the reference overall, but that a specific edge or component boundary remains distinguishable at all. A method can score
well on both metrics while still blurring together two adjacent parts that a designer needed
to tell apart, and conversely, a rendering could deviate noticeably in color or brightness from
the reference without costing the viewer any actual geometric information. Neither MSE nor
FLIP is designed to notice this distinction, since both operate on raw or perceptually-weighted
pixel differences rather than on whether discrete, task-relevant features remain separable in
the final image. This suggests that CAD-specific quality evaluation may ultimately need
metrics that more directly measure feature discriminability and edge preservation rather than
overall perceptual or numerical fidelity.

\section{VISIBILITY ENHANCEMENT FOR CAD}
\label{sec:important-parts}
\begin{figure}[!htb]
    \centering
    \includegraphics[width=1\linewidth]{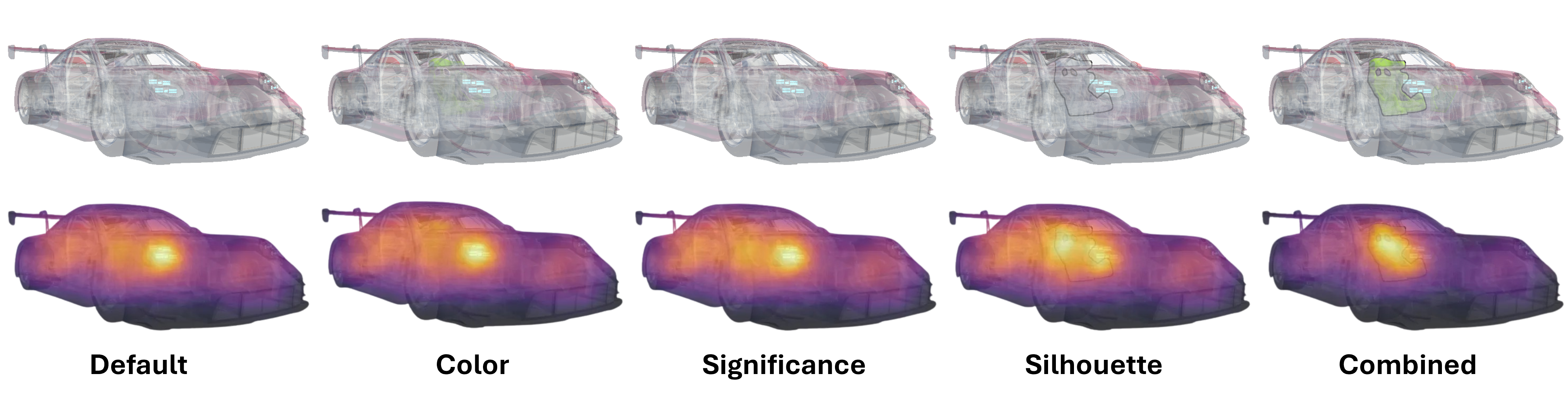}
    \caption{CAD model transparency enhancements applied to a complex car assembly (top) and corresponding saliency maps \cite{KRONER2020261} (bottom). From left to right: no enhancements, color highlighting, significance, silhouette, and combined enhancements. In complex scenarios, a combination of enhancements is required to correctly emphasize the selected area.}
    \label{fig:enhancementtypes}
\end{figure}

In this section we explore different alternatives of representing and visualizing CAD parts that we cannot afford to miss when rendering one or multiple CAD models with transparency.

These distinguish two broad types of approaches: 
\begin{itemize}
\item Data that is up to the designer to correctly incorporate in the CAD models: using colors, transparency values or a user defined field in the CAD model that determines the significance value of the specific part. Then algorithms are used to correctly render these semantics to conform to the designer intent.

\item System provided services: Employ object silhouettes, define at system level the importance objects and render them differently and finally alleviate z-fighting. These are provided by the system without any user-designer input at the CAD model. 
\end{itemize}

\subsection{CAD Model Driven Transparency Rendering}

\begin{figure}[!h]
    \centering
    \includegraphics[width=1\linewidth]{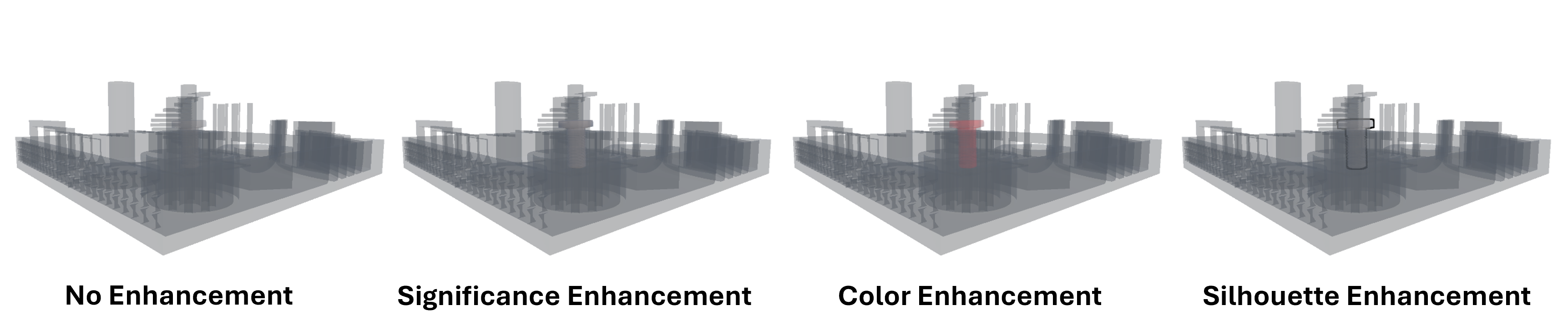}
    \caption{Different types of enhancements applied to a CAD use case.}
    \label{fig:enhancements}
\end{figure}

The user sets colors and transparency values to CAD models that emphasize important parts. For example, the user may set blue and red for certain parts considered important and light-gray for all other parts. Likewise, the user may set low transparency for important parts and high transparency for the rest. Or a combination of the above.
 
Alternatively, the user may set a per-part \emph{significance} multiplier that automatically increases opacity. A significance of $0$ leaves opacity unchanged; a significance of $1$ increases opacity by 100\% (i.e., doubles it), subject to the upper bound of full opacity. Formally, given a fragment's authored opacity $\alpha$ and significance $s \in [0, s_{max}]$,
\begin{equation}
\alpha' = \min\left(1,\ \alpha \cdot (1 + s)\right).
\label{eq:significance}
\end{equation}
 
We define significance as an \textit{opacity multiplier} rather than a technique-specific parameter for two reasons. First, CAD materials are predominantly assigned a single, uniform opacity per part rather than a spatially-varying alpha texture, so a scalar multiplier suffices for the overwhelming majority of CAD use cases. Second, and more importantly, every OIT method we survey ultimately resolves visibility through some function of per-fragment opacity, whether as an explicit blend weight, an over or under operator compositing term, or an implicit feature used by a neural network. Opacity is thus the one quantity common to all categories, allowing a single authoring-level concept to propagate across every method we evaluate, differing only in \emph{how} it propagates.
 
\paragraph{Direct propagation (blended-approximate methods).} In WBOIT, every fragment contributes through a single continuous accumulation step weighted by a function of opacity and depth \cite{McGuire2013}. Since this accumulation has no notion of discrete layers or rank, applying Eq.~\ref{eq:significance} to opacity prior to weighting is sufficient: the boosted opacity simultaneously increases visual contribution and accumulation weight, with no further changes to compositing or pass structure required.
 
\paragraph{Sort-biased propagation (buffer-based methods).} Exact and layer-based methods (k-buffer, A-buffer, and approximate methods such as DFAOIT) maintain only a bounded number of \emph{k} nearest fragments per pixel. Fragments that overflow this budget are either discarded or merged into a \emph{tail} term approximated through simple blending or neural networks, like DFAOIT \cite{tsopouridis2024dfa, tsopouridis2026starntspatiotemporalaccelerationrealtime}. Here, direct propagation alone is insufficient: scaling opacity changes how a fragment is composited \emph{within} the bucket it occupies, but not \emph{whether it is inserted} into the \emph{k}-buffer at all. This means that if a fragment spills into the tail, the foremost, exactly-sorted fragments always overshadow it, diminishing the effect of our direct opacity propagation.
 
We address this with a bounded perturbation of the depth value used for sorting. For depth $z$ and significance $s$, we define a biased depth
\begin{equation}
z' = \max(0,\ z - s \cdot \beta),
\end{equation}
where $\beta$ bounds the maximum insertion-order displacement. $z'$ is used \emph{only} for the depth-sort comparison, in essence promoting this fragment ahead of the tail and into the \emph{k}-buffer.
 
We implemented this for DFAOITs second depth-peeling pass: the insertion test uses $z'$, while eligibility (whether a fragment is farther than the first peeled fragment) continues to use unmodified $z$, so significance cannot promote a fragment nearer than the first peeled layer, preserving mostly correct fragment ordering. At $s=0$, $z'=z$ identically, preserving backward compatibility. We argue this pattern generalizes to any method whose behavior depends on a depth-sorted fragment list rather than continuous accumulation, notably k-buffer and A-buffer.

\begin{table}[h]
\centering
\caption{Significance propagation pattern by OIT category.}
\label{tab:significance-propagation}
\begin{tabular}{|l|l|l|}
\hline
\textbf{Category} & \textbf{Representative} & \textbf{Propagation} \\
\hline
Object-sorted & ODT & Direct \\
\hline
Exact / Buffer-based & A-buffer, k-buffer & Sort-biased \\
\hline
Blended-approximate & WBOIT & Direct \\
\hline
Hybrid/Neural Approximate & DFA & Direct + Sort-biased \\
\hline
\end{tabular}
\end{table}


\subsection{Silhouettes}
\label{sec:silhouettes}

In complex CAD assemblies, a selected or important part may be partially or fully occluded by surrounding transparent geometry, making it difficult to locate visually even when rendered with elevated significance. Silhouette-based highlighting addresses this by drawing a visible outline around the selected part's projected boundary on screen, providing a stable spatial cue regardless of occlusion.

\paragraph{Outline rendering.}
We render silhouettes using a two-pass screen-space mask approach. In a \emph{mask pass}, the selected part's geometry is re-drawn into a dedicated single-channel render target, writing a flat value of 1 at every covered pixel while respecting the existing camera depth buffer so that the silhouette is correctly occluded by opaque geometry in front of the part. The mask pass simultaneously writes the selected part's depth $z_{selected}$ into a second render target for use in transmittance estimation. A subsequent fullscreen \emph{composite pass} detects the mask boundary by sampling eight neighbors within a user-set pixel radius, pixels outside the silhouette that have at least one silhouette neighbor are classified as edge pixels and rendered with the outline color. This approach is structurally orthogonal to the underlying OIT method: it operates entirely on the resolved color buffer and does not interact with the OIT accumulation or compositing passes, and can therefore be composed with any OIT technique without modification to the OIT pipeline.

\paragraph{Transmittance-attenuated outlines.}
A flat outline of constant opacity is visually misleading when the selected part is heavily occluded by transparent geometry. We attenuate the outline opacity by an \textit{estimate of the transmittance} of the transparent geometry in front of the selected object, so that the outline fades naturally as occlusion increases, while remaining visible through a configurable minimum floor $v_{min}$.

The ideal attenuation factor is the front-only transmittance:
\begin{equation}
T_{front}(z_{selected}) = \prod_{i:\ z_i < z_{selected}} (1 - \alpha_i),
\label{eq:front-transmittance-ideal}
\end{equation}
i.e., the product of transmittances only over fragments in front of the selected part. No OIT method provides this directly, but every method decomposes fragments into two groups: a set of \emph{stored layers} represented exactly, and an \emph{aggregate tail} that collapses the remaining fragments into a compact representation. This motivates a unified estimator:
\begin{equation}
\hat{T}_{front}(z) = T_{layers}(z) \cdot T_{tail}^{\ f(z)},
\label{eq:front-transmittance-est}
\end{equation}
where $T_{layers}(z) = \prod_{i \in \text{stored},\ z_i < z}(1-\alpha_i)$ is the exact transmittance from stored layers shallower than $z_{selected}$, $T_{tail}$ is the aggregate tail transmittance, and $f(z) \in (0,1)$ is the estimated fraction of tail fragments in front of $z$. The latter is computed via a sigmoid centered on the mean linear depth $\bar{z}$ of all transparent fragments at the outline pixel:
\begin{equation}
f(z) = \frac{1}{1 + e^{-k(z - \bar{z})}},
\label{eq:sigmoid}
\end{equation}
where $k$ controls the sharpness of the transition. The mean depth $\bar{z} = \frac{1}{N}\sum_i z_i$ is obtained from a per-pixel depth sum and fragment count accumulated additively during the OIT pass at negligible cost. The intuition behind Eq.~\ref{eq:sigmoid} is straightforward: if the selected part is shallower than the average transparent fragment ($z_{selected} < \bar{z}$), most of the tail is likely \emph{behind} it and $f \to 0$, so the tail contributes little to front-only attenuation. Conversely, if the selected part is deeper than average ($z_{selected} > \bar{z}$), most of the tail is likely \emph{in front} and $f \to 1$, applying the full tail transmittance. The final attenuated outline opacity is:

\begin{equation}
\alpha_{outline} = \text{lerp}\!\left(1,\ \max\!\left(\hat{T}_{front},\ v_{min}\right),\ \lambda\right),
\label{eq:outline-attenuation}
\end{equation}
where $\lambda \in [0,1]$ is a global strength parameter and $v_{min}$ ensures the outline remains visible.

For exact methods (A-buffer), $T_{tail} = 1$ and $\hat{T}_{front}$ is exact. For k-buffer and hybrid methods, stored layers contribute $T_{layers}(z)$ exactly while the tail uses the sigmoid estimate. For WBOIT, $T_{layers}(z) = 1$ and the formula reduces to $\hat{T}_{front} = R^{f(z)}$ using the existing revealage buffer. MBOIT \cite{Munstermann2018} can evaluate $T_{front}(z_{selected})$ directly from its moment representation without the sigmoid. For DFAOIT, the two peeled layers contribute $T_{layers}(z)$ exactly and the tail is approximated as $T_{tail} \approx (1-\bar{\alpha})^N$ from statistics already available in the feature extraction pass. The main source of approximation in all non-exact cases is the sigmoid estimate of $f(z)$, which assumes that tail fragments are distributed in depth around $\bar{z}$. This fails when transparent geometry is heavily concentrated on one side of the selected object, causing over- or under-attenuation accordingly. In practice this is rarely a concern for CAD: assemblies consist of many distinct parts at varying depths, and transparency is typically assigned uniformly across part families rather than concentrated at a single depth layer, producing a spread depth distribution that is precisely the regime where the sigmoid estimate is most reliable. A more accurate alternative would require depth-stratified transmittance buffers.

\subsection{Z-Fighting Awareness}
CAD assemblies frequently contain overlapping surfaces: duplicated faces from boundary-representation exports, or thin parts modeled with zero clearance against an adjacent component. For methods that resolve visibility through a depth comparison, such coincident fragments cannot be reliably ordered, as floating-point depth values that are equal, or differ by less than the precision of the stored depth buffer, cause the comparison's outcome to depend on draw order or GPU scheduling rather than the geometry itself. The visible symptom is \textbf{\textit{flickering}} as the fragment selected for a given layer or pixel changes unpredictably across frames, an artifact long recognized in the depth-peeling literature as Z-fighting~\cite{vasilakiszfighting2011}.
 
We address this by extending the depth comparison used for fragment ordering with a secondary key, so that ties in depth are resolved deterministically rather than arbitrarily. For depth $z$ and a per-fragment identifier $id$ (the built-in primitive or object identifier), we pack both into a single value
\begin{equation}
\kappa(z, id) = \lfloor z \cdot 2^{b_z} \rfloor \cdot 2^{b_{id}} + (id \bmod 2^{b_{id}}),
\end{equation}
with $b_z + b_{id} = 24$ so that $\kappa$ remains exactly representable in a 32-bit float, as the default depth-buffer size is usually 24-bit. Ordering by $\kappa$ is equivalent to ordering by depth first and primitive identifier second, with no epsilon term required: two fragments compare equal under $\kappa$ only if they are quantized to the same depth bucket \emph{and} share the same identifier.

This packed key can be substituted for the raw depth comparison in any buffer-based OIT method that orders or sorts fragments by depth: in depth-peeling methods, $\kappa$ replaces the comparison used to determine the next fragment to peel, in k-buffer and A-buffer, $\kappa$ replaces the key used for insertion into the per-pixel fragment list. In every case, the only requirement is that the method utilizes per-fragment depth comparisons, which the packed key transparently replaces without otherwise altering the method's structure.
 
\section{DISCUSSION, CHALLENGES AND CONCLUSIONS}
\label{sec:conclusions}
This paper underscores the central role of transparency in CAD and reviews exact, approximate, hybrid, and neural OIT techniques through the lens of performance–quality trade-offs. We further present complementary visibility tools that preserve correct compositing while improving readability: significance-driven emphasis, transmittance-aware silhouettes, and z-fighting-robust ordering. Next, we provide a practical roadmap for selecting transparency methods across CAD tasks and hardware tiers, and conclude with open research directions.

\subsection{Which Transparency Method is Appropriate for my CAD platform?}
Choosing a transparency rendering method for CAD software involves balancing rendering quality against hardware requirements and performance cost. The main findings are:

\begin{itemize}[leftmargin=*]
    \item \textit{A-buffer implementations} are appropriate for a medium number of
    transparent layers (e.g., 20-30) and require high-end workstations and GPUs.
    They offer high (exact) quality, but memory usage scales with depth complexity.

    \item \textit{MBOIT and k-buffer/hybrid} methods are appropriate for high-end
    workstations, particularly their neural variants, which further improve
    reconstruction quality and can be adapted for CAD use cases.

    \item \textit{X-ray, screen-door, and ODT} techniques have unpredictable visibility
    and produce significant artifacts, though they carry zero OIT overhead.

    \item \textit{WBOIT} is very fast and lightweight, but suffers from low visibility
    (blurring), making it a viable option mainly for mobile devices and legacy GPUs.

    \item \textit{DFAOIT} provides high quality with fast, fixed-cost rendering,
    especially in its accelerated form (STAR-NT), and works well on mobile and
    legacy hardware.
\end{itemize}

Overall, high-end workstations can favor A-buffer, k-buffer/hybrid, or MBOIT/DFAOIT methods
(especially their neural variants) for maximum quality, while mobile or legacy hardware
is better served by WBOIT or DFAOIT, with DFAOIT (and its STAR-NT \cite{tsopouridis2026starntspatiotemporalaccelerationrealtime} acceleration) offering
the best quality-to-performance tradeoff on constrained hardware.

\subsection{Open Research Problems}

Robust, real-time transparency for Constructive Solid Geometry (CSG) remains open, especially when CSG trees are rendered alongside conventional B-rep/mesh objects. While multi-fragment approaches can render CSG without conversion to B-reps \cite{rossignac}, achieving correct per-pixel compositing, stable depth ordering, and competitive performance under high depth complexity is still challenging.

Delivering accurate, order-independent transparency with task-driven highlighting on constrained hardware (legacy desktops, mobile, and AR devices) is critical for maintenance and inspection in large industrial settings (e.g., plants, aerospace, naval). The research need is for fixed-cost methods that preserve edges and silhouettes, remain temporally stable, and degrade gracefully under tight compute, memory, and latency budgets.

Standard rendering metrics (FLIP, MSE, SSIM, LPIPS) do not capture whether parts of a complex CAD model are perceived in the correct order. In addition, we should verify that parts designated as important or of interest are visually detectable in the correct sequence. Saliency-based metrics, such as those explored experimentally in Section \ref{sec:important-parts}, may be more appropriate for this purpose.

A further avenue is extending the illustration buffer concept \cite{illustration-buffer} to transparency-aware, CAD-centric visualization. This includes integrating importance cues, silhouettes, and metadata/PMI-product manufacturing information- overlays into a unified, real-time pipeline that respects correct blending while supporting interactive exploration of complex solid modeling tasks.


\bigskip
\orcid{Grigoris Tsopouridis}{0000-0002-4137-0986}
\orcid{Ioannis Fudos}{0000-0001-8033-5481}


\referenceSection
\bibliographystyle{CADA}
\bibliography{CADandA_Paper_Template}

\bigskip
\end{document}